\documentclass[12pt,psfig]{article}
\usepackage{amssymb}
\usepackage{amsmath,amsthm}
\usepackage{amsfonts}
\usepackage{graphicx}
\usepackage{graphics}
\numberwithin{equation}{section}

\usepackage{hyperref}

\begin{document}
\begin{center}\Large\textbf{ Moving Fractional Branes with Background Fields:
Interaction and Tachyon Condensation}
\end{center}
\vspace{0.75cm}
\begin{center}{\large  Maryam Saidy-Sarjoubi and \large Davoud
Kamani}\end{center}
\begin{center}
\textsl{\small{Physics Department, Amirkabir University of
Technology (Tehran Polytechnic)\\
P.O.Box: 15875-4413, Tehran, Iran\\
 e-mails: kamani@aut.ac.ir , mrymsaidy@aut.ac.ir\\
}}
\end{center}
\vspace{0.5cm}
\begin{abstract}

We calculate the
bosonic boundary state corresponding to a moving fractional
D$p$-brane in a partially orbifoldized spacetime 
$\mathbb{R}^{1, d-5} \times\mathbb{C}^{2}/\mathbb{Z}_{2}$ 
in the presence of the Kalb-Ramond field, the $U(1)$ gauge potential
and the tachyon field. Using this boundary state we
obtain interaction amplitude of two parallel moving D$p$-branes with 
the above background
fields. Various properties of the interaction will be investigated. 
Besides, we study effects of the tachyon condensation on 
a moving fractional D$p$-brane with the above background fields
through the boundary state formalism.

\end{abstract}


\newpage
\section{Introduction}
\hspace{0.5cm} 

Boundary states, that first appeared in the
literature \cite{bon}, \cite{bon2}, have a central role in string theory
and D-branes. They
have been used to study D-brane properties and their interactions\cite{hus}, 
\cite{bergf}.
Precisely, interaction between two D-branes can be described in two
different ways: the open and closed string channels. In the open
string channel the interaction amplitude is given by the one-loop
diagram of the open string, stretched between two D-branes,
\cite{pol1}, \cite{pol0}, \cite{bach}, hence it is a quantum process. 
In the closed string channel one
can describe the interaction between the branes via the 
tree-level exchange of a closed string that is emitted from the
first brane then propagates toward the second one and is absorbed
there \cite{cal0}, \cite{green1}, \cite{greenl}, \cite{green2}, 
thus it is a classical process. In this 
approach each brane couples to all closed
string states via the boundary state corresponding to the brane.
This is because of the boundary state encodes all properties of the D-branes.
However, these two approaches of
interaction between the branes are equivalent and this
equivalence is called the open/closed string duality \cite{khor}.

On the other hand, the D-branes with nonzero background internal fields have
shown several interesting properties 
\cite{fra}, \cite{calli}, \cite{guk}, \cite{li}, 
\cite{kit}, \cite{bach0}, \cite{doc1}. Therefore,
the boundary state formalism for various setups
of D-branes in the presence of background fields
such as $B_{\mu\nu}$, the $U(1)$ gauge field and tachyon field in
the compact spacetime have been investigated.
However, among the various setups with two D-branes the systems 
with fractional branes have some interesting behaviors
\cite{frac0}, \cite{dog1}, \cite{frac1}, \cite{frac2}, \cite{frac3}, \cite{dive}. 
For example,
in \cite{dive} the gauge/gravity correspondence is derived from the open/closed
string duality for a system of fractional branes.

Another important issue concerning the D-branes
is the stability of them. 
The stability (instability) of D-branes can be investigated via the 
open string tachyon condensation\cite{cn2}, \cite{cn3}.
This condensation usually leads to the instability and collapse of 
the D-branes. That is, an unstable D-brane decays into a lower dimensional
unstable D-brane as an intermediate state, and finally to the closed 
string vacuum. These concepts have been studied by various methods \cite{cn4},
\cite{cn5}, \cite{cn6}, \cite{cn7}.
Since the boundary state completely 
comprises all properties of the brane it can be 
used to investigate the time evolution of the brane 
through the tachyon condensation process
\cite{cn8}, \cite{cn9}, \cite{cn10}, \cite{cn11}.

In this paper we use the boundary state method to obtain 
the interaction amplitude between two parallel moving fractional
D$p$-branes in a factorizable spacetime with the orbifold 
structure  
$\mathbb{R}^{1, d-5} \times\mathbb{C}^{2}/\mathbb{Z}_{2}$. 
We shall consider
the Kalb-Ramond field $B_{\mu\nu}$, the $U(1)$ gauge potential 
and the tachyon field on the worldvolumes of the branes.
In addition, the branes are moving along a common axis 
which is perpendicular to both of them. Thus, in this 
setup the generality of the system has been exerted,
which drastically affects the interaction of the branes.  
We shall also study long-time
behavior of the interaction amplitude. Besides, we shall 
investigate effects of tachyon condensation on the  
stability of a moving fractional D-branes. We shall 
observe that condensation of the tachyon drastically reduces 
the dimensions of such branes.

The paper is organized as follows. In Sec. 2, we compute
the boundary state associated with a 
moving fractional D$p$-brane 
with various background fields. In Sec. 3, we find the interaction
amplitude of two parallel such branes, 
and its behavior for large distances of the branes. In Sec. 4, 
we examine a moving fractional D$p$-brane with various fields
under the experience of the tachyon condensation.
Section 5 is devoted to the conclusions.
\section{The boundary state of D$p$-brane}
\hspace{0.5cm}

Consider a fractional D$p$-brane which lives in the $d$-dimensional
spacetime, including the orbifold
$\mathbb{C}^{2}/\mathbb{Z}_{2}$, where the $\mathbb{Z}_2$ group
acts on the coordinates $\{x^a|a= d-4, d-3, d-2, d-1\}$. This
orbifold is noncompact, so its fixed points are located at $x^a
=0$. The D$p$-brane is stuck at these fixed points.

We start with the following sigma-model action for the closed string
\begin{eqnarray}
ýSý=ý&-&ýý\frac{1}{ý4\pi\alpha'ý}ýý\int_{\Sigma} d^2\sigmaý
\left(\sqrt{-h} h^{ab}G_{\mu\nu} ý\partialý_{a}X^\mu
ýý\partial_b X^{\nu} +\epsilon^{ab}B_{\mu\nu}\partial_a ýX^{\mu}
ý\partial_b ýX^{\nu}\right)
\nonumber\\
&+&\frac{1}{2\pi\alpha'}\int_{\partial\Sigma}d\sigma\left(A_{\alpha}
\partial_{\sigma}X^{\alpha}
+\frac{i}{2}U_{\alpha \beta}X^{\alpha}X^{\beta}\right)~,
\end{eqnarray}
where the set $\{x^\alpha | \alpha = 0,1, \cdot\cdot\cdot,p\}$
represents the brane directions,
$\Sigma$ indicates the worldsheet of the closed string, and
$\partial\Sigma $ is the boundary of it. The metrics of the
worldsheet and the $d$-dimensional spacetime are $h_{ab}$ and
$G_{\mu\nu}$, respectively. For simplifying the equations we
select the Kalb-Ramond field $B_{\mu\nu}$ to be constant and 
$G_{\mu\nu}=\eta_{\mu\nu} ={\rm
diag}(-1,1,\cdot\cdot\cdot,1)$. The tachyon profile is chosen as
$T(X) = \frac{i}{4\pi \alpha'}U_{\alpha \beta}X^{\alpha}X^{\beta}$ with constant
symmetric matrix $U_{\mu\nu}$. We chose the tachyon
field only in the worldvolume of the D$p$-brane.
For the $U(1)$ gauge potential $A_{\alpha}$,
which lives on the worldvolume of the
brane, we consider the gauge 
$A_{\alpha}=-\frac{1}{2}F_{\alpha \beta }X^{\beta}$ where 
the field strength is constant.
Note that the gauge and tachyon
fields are in the open string spectrum, and hence 
their open string state counterparts adhere to the brane.

Vanishing variation of this action defines the following 
boundary state equations for closed string
\begin{eqnarray}
&~&\left(\partial_{\tau}X^{\alpha}
+\mathcal{F}^{\alpha}_{~~\beta}\partial_\sigma
X^{\beta}-iU^{\alpha}_{~\beta}
X^\beta\right)_{\tau=0}|B_x\rangle=0~,
\nonumber\\
&~&\left(X^I-y^I\right)_{\tau=0}|B_x\rangle=0~,
\end{eqnarray}
where the coordinates $\{x^I|I = p+1, \cdot\cdot\cdot,d-1\}$ 
refer to the directions
perpendicular to the brane worldvolume and the parameters 
$\{y^I\}$ specify the
location of the brane. For more simplification we assumed that the mixed
elements $B^\alpha_{~~I}$ are zero. The total field strength
possesses the definition
$\mathcal{F}_{\alpha\beta}=F_{\alpha\beta} -B_{\alpha\beta}$.

Note that because the brane is stuck at the orbifold fixed points,
presence of the orbifold directions puts some prominent constraints on its
dimension and motion. In the $d$-dimensional spacetime 
the brane can possesses the maximum dimension $d-5$. Besides, 
along the orbifoldized directions it can not move.
Therefore, for adding a velocity to the brane along the perpendicular
directions $\{x^i|i=p+1,\cdot\cdot\cdot,d-5\}$ we apply a
boost on the Eqs. (2.2),
\begin{eqnarray}
&~&[\partial_{\tau}(X^0-v^i X^i)+\mathcal{F}^{0}_{~~{\bar
\alpha}}
\partial_\sigma
X^{{\bar \alpha}}-iU^0_{~~0}\gamma^2 (X^0-v^i X^i)-iU^0_{~~{\bar
\alpha}} X^{\bar \alpha}]_{\tau=0}|B_x\rangle=0~,
\nonumber\\
\nonumber
&~&[\partial_{\tau}X^{\bar{\alpha}}
+\gamma^2 \mathcal{F}^{\bar{\alpha}}_{~~0}
\partial_\sigma (X^0-v^i X^i)
+\mathcal{F}^{\bar{\alpha}}_{~~{\bar \beta}}
\partial_\sigma X^{\bar \beta}-iU^{\bar \alpha}_{~~0}\gamma^2(X^0-v^i X^i)
-iU^{\bar{\alpha}}_{~{\bar \beta}} X^{\bar
\beta}]_{\tau=0}|B_x\rangle=0~,
\nonumber\\
\nonumber
&~&[X^i-v^i X^{0}-y^i]_{\tau=0}|B_x\rangle=0~,
\nonumber\\
&~&[X^a-y^a]_{\tau=0}|B_x\rangle=0~,
\end{eqnarray}
where $\gamma = 1/ \sqrt{ 1 - v^i v^i}$, 
the set $\{x^{\bar \alpha}\}$ shows the directions
of the brane, and the set $\{x^i\}$ indicates the directions 
perpendicular to its worldvolume except the orbifoldized directions. 
Since the branes are stuck at the
orbifold fixed points we have $y^{a}=0$. 

The mode expansion of the closed string coordinates along the
non-orbifold directions $x^\alpha$ and $x^i$ has the feature 
\begin{equation}
X^\lambda(\sigma,\tau)=x^\lambda+2\alpha'p^\lambda
\tau+\frac{i}{2}\sqrt{2\alpha'}\sum_{m\neq0}\frac{1}{m}
\left(\alpha_m^{\lambda}e^{-2im(\tau- \sigma)}+\tilde{\alpha}_m^\lambda
e^{-2im(\tau+\sigma)}\right)~,\;\lambda \in \{\alpha , i\},
\end{equation}
and for the orbifold directions takes the form  
\begin{eqnarray}
X^a(\sigma,\tau)=\frac{i}{2}\sqrt{2\alpha'}\sum_{r\in\mathbb{Z}+\frac{1}{2}}
\frac{1}{r}\left(\alpha_r^{a}e^{-2ir(\tau- \sigma)}+\tilde{\alpha}_r^a
e^{-2ir(\tau+\sigma)}\right),
\end{eqnarray}
Now for simplification we suppose $U_{0\alpha}=U_{\alpha0}=0$.
Using the above mode expansions the boundary state equations (2.3)
can be written in terms of the string oscillators and zero-modes 
\begin{eqnarray}
&~&[\alpha^{0}_m-v^{i}\alpha^{i}_m-
\mathcal{F}^{0}_{~~\bar{\alpha}}
\alpha^{\bar{\alpha}}_m +
\tilde{\alpha}^{0}_{-m}-v^{i}\tilde{\alpha}^{i}_{-m}+
\mathcal{F}^{0}_{~~\bar{\alpha}}
\tilde{\alpha}^{\bar{\alpha}}_{-m}
]|B_{\rm osc} \rangle=0~,
\nonumber\\
\nonumber
&~&[\alpha^{\bar{\alpha}}_m-\gamma^2 \mathcal{F}^{\bar{\alpha}}_{~~0}
(\alpha^{0}_m-v^{i}\alpha^{i}_{m})-\mathcal{F}^{\bar{\alpha}}_{~~\bar\beta}
\alpha^{\bar\beta}_m +\frac{1}{2m}U^{\bar{\alpha}}_{~~\bar\beta}
\alpha^{\bar\beta}_m
\nonumber\\
&~&+\tilde{\alpha}^{\bar{\alpha}}_{-m}+\gamma^2
\mathcal{F}^{\bar{\alpha}}_{~~0}
(\tilde\alpha^{0}_{-m}-v^{i}\tilde\alpha^{i}_{-m})+
\mathcal{F}^{\bar{\alpha}}_{~~\bar\beta}
\tilde{\alpha}^{\bar\beta}_{-m}-\frac{1}{2m}U^{\bar{\alpha}}_{~~\bar\beta}
\tilde{\alpha}^{\bar\beta}_{-m}]|B_{\rm osc} \rangle=0~,
\nonumber\\
\nonumber
&~&[\alpha^{i}_m-v^{i}\alpha^{0}_m-\tilde\alpha^{i}_{-m}+v^{i}\tilde
\alpha^{0}_{-m})]|B_{\rm osc} \rangle=0~,
\nonumber\\
&~&(\alpha^{a}_r-\tilde\alpha^{a}_{-r})|B_{\rm osc} \rangle=0~,
\end{eqnarray}

\begin{eqnarray}
&~&(\hat{p}^0-v^{i}\hat{p}^{i})
|B\rangle^{(0)}=0~,
\nonumber\\
\nonumber
&~&[2\alpha'\hat{p}^{\bar{\alpha}}-iU^{\bar{\alpha}}_{~~\bar\beta}
\hat{x}^{\bar\beta}]|B\rangle^{(0)}=0~,
\nonumber\\
\nonumber
&~&(\hat{p}^{i}-v^{i}\hat{p}^{0})|B\rangle^{(0)}=0~,
\nonumber\\
&~&[\hat{x}^{i}-v^{i}\hat{x}^0-y^{i}]|B\rangle^{(0)}=0~.
\end{eqnarray}
Note that we decomposed the boundary state as
$|B_x \rangle =|B_{\rm osc} \rangle \otimes |B \rangle^{(0)}$.
Since the closed string is emitted (absorbed) at the brane position 
$x^a=0$ the zero-mode 
equations don't have any contribution from $X^{a}$'s.
The second equation of (2.7), in terms of the eigenvalues, implies
the relation 
\begin{eqnarray}
p^{\bar \alpha} =\frac{i}{2\alpha'} 
U^{\bar \alpha}_{\;\;{\bar \beta}} x^{\bar \beta}.
\end{eqnarray}
Thus, in the brane volume 
the momentum of the emitted (absorbed) closed string 
depends on its center of mass
position. Thus, we deduce that the tachyon field inspires  
a peculiar potential on the closed string.
 
Using the coherent state method the oscillating part of the
boundary state possesses the solution 
\begin{eqnarray}
|B_{\rm osc}\rangle &=&\prod_{n=1}^{\infty}[\det{M_{(n)}}]^{-1}
\exp\left[{-\sum_{m=1}^{\infty}
\left(\frac{1}{m}\alpha_{-m}^{\lambda}S_{(m)\lambda\lambda'}
\tilde{\alpha}_{-m}^{\lambda'}\right)}\right]\nonumber\\
\nonumber\\
&\times&\exp\left[-\sum_{r=1/2}^{\infty}
\left(\frac{1}{r}\alpha_{-r}^{a}\tilde{\alpha}_{-r}^{a}\right)\right]
|0\rangle_\alpha|0\rangle_{\tilde{\alpha}}~,\label{aos}
\end{eqnarray}
where the infinite product comes from path
integral, and can be learned by the Refs. \cite{cal1}, \cite{cal2}.
Note that $\lambda , \lambda' \in \{\alpha , i\}$.
The matrix $S_{(m)}$ is defined as $S_{(m)}=M_{(m)}^{-1}N_{(m)}$
with 
\begin{eqnarray}
M_{(m)\lambda}^{0}&=&\gamma(\delta_{\lambda}^{0}-v^{i}\delta_{\lambda}^{i})-
\gamma\mathcal{F}^{0}_{~~\bar\alpha}\delta_{\lambda}^{\bar\alpha}~,
\nonumber\\
M_{(m)\lambda}^{\bar{\alpha}}&=&\delta_{\lambda}^{\bar{\alpha}}-\gamma^2
\mathcal{F}^{\bar{\alpha}}_{~~0}(\delta_{\lambda}^{0}-v^{i}
\delta^{i}_{\lambda})-(\mathcal{F}^{\bar{\alpha}}_{~~\bar\beta}
-\frac{1}{2m}
U^{\bar{\alpha}}_{~~\bar\beta})\delta^{\bar\beta}_{\lambda}~,
\nonumber\\
M_{(m)\lambda}^{i}&=&\delta_{\lambda}^{i}-v^{i}\delta_{\lambda}^{0}~.
\nonumber\\\nonumber
\\\nonumber
N_{(m)\lambda}^{0}&=&\gamma(\delta_{\lambda}^{0}-v^{i}\delta_{\lambda}^{i})+
\gamma\mathcal{F}^{0}_{~~\bar\alpha}\delta_{\lambda}^{\bar\alpha}~,
\nonumber\\
N_{(m)\lambda}^{\bar{\alpha}}&=&\delta_{\lambda}^{\bar{\alpha}}+\gamma^2
\mathcal{F}^{\bar{\alpha}}_{~~0}(\delta_{\lambda}^{0}-v^{i}
\delta^{i}_{\lambda})+(\mathcal{F}^{\bar{\alpha}}_{~~\bar\beta}
-\frac{1}{2m}
U^{\bar{\alpha}}_{~~\bar\beta})\delta^{\bar\beta}_{\lambda}~,
\nonumber\\
N_{(m)\lambda}^{i}&=&-\delta_{\lambda}^{i}+v^{i}\delta_{\lambda}^{0}~.
\end{eqnarray}
The Eq. (2.9) elaborates that a boundary state describes 
creation of all closed string states from vacuum, 
or equivalently
it represents a source for closed strings, emitted by the D-brane.

In fact, the coherent state method on the boundary state (2.9)
imposes the constraint $S_{(m)}S_{(-m)}^{T}=\mathbf{1}$,
which introduces some relations among the parameters
$\{v^{i},U_{{\bar \alpha}{\bar \beta}},\mathcal{F}_{\alpha\beta}\}$,
hence reduces the number of independent parameters.

The zero-mode part of the boundary state, i.e. the solution 
of Eqs. (2.7), is given by 
\begin{eqnarray}
|B\rangle^{(0)}=&~&\frac{T_p}{2\sqrt{\det(U/4\pi \alpha')}}
\int_{-\infty}^{\infty}\prod_{\lambda}
dp^{\lambda}\exp\left[-\alpha'(U^{-1})_{\bar\alpha\bar\beta}\;p^{\bar
\alpha}p^{\bar\beta}
\right]\nonumber\\
&~&\times \prod_i
\delta \left(\hat{x}^{i}-v^{i}\hat{x}^{0}-y^{i}\right)
\prod_{i}|p^{i}\rangle\prod_{\alpha}|p^{\alpha}\rangle~
\label{zer}.
\end{eqnarray}

The total boundary state associated with the D$p$-brane is
exhibited by the following direct product 
$$|B\rangle=|B_{\rm osc}\rangle \otimes|B\rangle^{(0)}
\otimes|B_{\rm gh}\rangle~,$$ 
where $|B_{\rm gh}\rangle$ is the boundary state of the 
anti-commuting ghosts
\begin{equation}
|B_{\rm gh}\rangle=\exp{\left[\sum_{m=1}^{\infty}(c_{-m}\tilde{b}_{-m}
-b_{-m} \tilde{c}_{-m})\right]}\frac{c_0+\tilde{c}_0}{2}
|q=1\rangle|\tilde{q}=1\rangle~.
\end{equation}
Since the ghost fields do not interact with the matter part, their
contribution to the boundary state is not affected by the orbifold
projection, the brane velocity and the background fields.
\section{Interaction of the D$p$-branes}

In this section we calculate the interaction amplitude between
two parallel-moving fractional D$p$-branes through the closed string exchange. 
For this, we
compute the overlap of the two boundary states via the closed
string propagator, i.e. $\mathcal{A}=\langle
B_1|D|B_2\rangle$, where $|B_1\rangle$ and $|B_2\rangle$
are the total boundary states corresponding to the branes, and  
$D$ is the closed string propagator which is accurately defined by
$$D=2\alpha'\int_{0}^{\infty}dt~e^{-tH_{\rm closed}}~.$$
The closed string Hamiltonian is sum of the Hamiltonians of
the matter part and ghost part. For the matter part there is
\begin{equation}
H_{\rm matter}=\alpha'p^{\lambda}p_{\lambda}
+2\left(\sum_{n=1}^{\infty}(\alpha_{-n}^{\lambda}
\alpha_{n\lambda}
+\tilde{\alpha}_{-n}^{\lambda}\tilde{\alpha}_{n\lambda})
+\sum_{r=1/2}^{\infty}
(\alpha_{-r}^{a}\alpha_{ra}
+\tilde{\alpha}_{-r}^{a}\tilde{\alpha}_{ra})\right)-\frac{d-4}{6}~.
\label{asd}
\end{equation}
The difference of the constant term with the conventional 
case is a consequence of the orbifold projection on the four directions.

For simplicity we suppose that the branes are moving
along the same alignment with the velocities $v^i_{1}$ and $v^i_{2}$. 
The result of the calculations reveals the following 
elegant interaction amplitude 
\begin{eqnarray}
\mathcal{A}&=&\frac{T_p^2\alpha'V_{\bar{\alpha}}}{2(2\pi)^{d-p-5}}
~\frac{\prod_{n=1}^{\infty}
\left[\det \left(M_{(n)1} M_{(n)2}\right)\right]^{-1}}
{\sqrt{\det{(U_1/4\pi \alpha')}
\det{(U_2/4\pi \alpha')}}}
\int_{0}^{\infty}dt\bigg{[}(\det \mathbf{A})^{-1/2}~e^{\frac{d-4}{6}t}
~\nonumber\\
\nonumber\\
&\times&
\Big{(}\sqrt{\frac{\pi}{\alpha' t}}\Big{)}^{d-p-5}
\exp\left( {-\frac{1}
{4\alpha't}\sum_{i}{\left(y_{1}^{i}
-y_{2}^{i}\right)^2}}\right)\bigg{]}
\nonumber\\
\nonumber\\ &\times& 
~\prod_{n=1}^\infty \bigg{(} 
\det[1-S^{T}_{(n)1}S_{(n)2}e^{-4nt}]^{-1}~
(1- e^{-4nt})^{2}(1- e^{-2(2n-1)t})^{-4}\bigg{)}\bigg\},
\label{tg}
\end{eqnarray}
where $V_{\bar{\alpha}}$ is the common volume of the branes, and 
\begin{eqnarray}
\mathbf{A}_{\bar\alpha\bar\beta}=2\alpha't\delta_{\bar\alpha\bar\beta}
-2\alpha'\big{[}
(U^{-1}_{1})_{\bar\alpha\bar\beta}
-(U^{-1}_{2})_{\bar\alpha\bar\beta}\big{]}.
\label{aa}
\end{eqnarray} 
In the second line the exponential term indicates a damping
factor concerning to the distance of the branes. In the last 
line the determinant part is contribution of the 
oscillators of the non-orbifoldy directions,
while advent of $\prod_{n=1}^\infty(1- e^{-4nt})^2$ 
is due to the conformal ghosts.
The overall factor behind the integral, which depends on the parameters
of the system, clarifies a portion of the interaction strength.
\subsection{Interaction of the distant branes}

In any interaction theory, 
behavior of interaction amplitude, after an enough long time, 
gives a trusty 
long-range forces of the theory. On the other hand, 
for the distant branes the massless closed string states 
possess a considerable contribution on the interaction, 
while the contribution of all massive states, except the tachyon 
state, are damped.

The orbifold projection specifies some new effects on the 
large distance amplitude. This interaction  
is constructed via the limit $t\rightarrow \infty$
of the oscillating part of the general amplitude (3.2). Therefore, the
contribution of the graviton, Kalb-Ramond, dilaton and 
tachyon states on the interaction in the 26-dimensional
spacetime is determined by  
\begin{eqnarray}
\mathcal{A}_0&=&\frac{T_p^2\alpha'V_{\bar{\alpha}}}{2(2\pi)^{d-p-5}}
~\frac{\prod_{n=1}^{\infty}\left[\det
\left(M_{(n)1} M_{(n)2}\right)\right]^{-1}}{\sqrt{\det{(U_1/4\pi \alpha')}
\det{(U_2/4\pi \alpha')}}}~\nonumber\\
\nonumber\\
&\times&\int^{\infty} dt
\Big{(}\sqrt{\frac{\pi}{\alpha' t}}\Big{)}^{d-p-5}
\exp{\left(-\frac{1}
{4\alpha't}\sum_i
\left(y_{1}^{i}-y_{2}^{i}\right)^2\right)}
\nonumber\\
\nonumber\\
~&\times&(\det \mathbf{A})^{-1/2}~\left(e^{11t/3}+{\rm Tr}
(S^{T}_{(n=1)1} S_{(n=1)2})e^{-t/3}\right).
\end{eqnarray}
We applied the limit only on the third line of 
Eq. (3.2). This is due to the fact that the other factors do not
originate from the exchange of the massless 
and tachyon states. For example,
the exponential factor is related to the position of the branes. 
Appearance of the divergent part  
is a subsequent of the exchange of the closed string tachyon,
due to its negative mass squared. At the limit
$t\rightarrow\infty$ the second factor in the last line 
vanishes. This demonstrates that the massless states,
i.e. the gravitation, dilaton and Kalb-Ramond, 
prominently do not possess any contribution in the 
long distant interaction. In other words,
orbifold projection quenches the long 
range force of the twisted sector. 
More precisely, this projection manipulated  
the zero point energy of the Hamiltonian, hence, this 
result was created. Note that the untwisted
sector of the theory possesses the long-range force. 
Hence, the total amplitude which
comes from the both twisted and untwisted sectors 
contains a non-vanishing long-range force.
Note that the massless states, similar to the massive ones, 
for usual distances of the branes contribute to the interaction.

\section{Instability of a D$p$-brane under the tachyon condensation}

One of the main important aspects of studying the D-branes 
is determining their stability or instability, which drastically leads to finding 
the time evolution of them. Generally, adding the tachyonic mode 
of the open string spectrum to a single D-brane or to a system of D-branes 
usually makes them unstable. 
This phenomenon is known as tachyon condensation \cite{cn2}, \cite{cn3}. 
During this process the dimension of the brane 
is consecutively reduced and at the end we receive only closed strings. 
In this section we examine the behavior of our D$p$-brane 
under the experience of the condensation of the tachyon. Our aim
is to see the effects of the fractionality, transverse motion and background 
fields on the stability of the brane.

Tachyon condensation occurs when some of
the elements of the tachyon matrix become infinity. 
We exhibit the condensation 
via the limit $U_{pp}\rightarrow \infty$.
To obtain evolution of the D$p$-brane 
we apply this limit on the corresponding 
boundary state. At first we observe that 
since there is no tachyon matrix element in the 
orbifold part of the boundary state,
the condensation of tachyon has no effect on this part.  
This elaborates that fractionality of the brane on its instability 
is inactive.

The limit $U_{pp}\rightarrow \infty$ implies that 
\begin{equation}
\lim_{U_{pp}\rightarrow \infty}(U^{-1})_{p\bar{\alpha}}=
\lim_{U_{pp}\rightarrow \infty}(U^{-1})_{\bar{\alpha}p}=0.
\end{equation}
Therefore, the dimensional reduction on the exponential factor 
of Eq. (\ref{zer}) takes place, i.e. the matrix
$(U^{-1})_{\alpha'\beta'}$ with $\alpha',~\beta' \neq p$,
which is $(p-1)\times (p-1)$, appears. 

The prefactor of the total boundary state is
\begin{equation}
\frac{T_p}{2}\frac{\prod_{n=1}^{\infty}\left[\det
M_{(n)}\right]^{-1}}{\sqrt{\det{(U/4\pi \alpha')}}}~.\label{infty}
\end{equation}
Now we find evolution of this factor after condensation of the tachyon.
Thus, we have
$$\lim_{U_{pp}\rightarrow \infty}\det U_{p\times p}
=U_{pp}\;\det{\tilde{U}}_{(p-1)\times(p-1)},$$
where the matrix $\tilde{U}$ completely is similar to $U$ 
without the last row and the last column. 
In the same way, for the matrix $M_{(n)}$ we acquire  
$$\lim_{U_{pp}\rightarrow \infty}\det 
\left( M_{(n)}\right)_{(d-4)\times (d-4)}=
\frac{1}{2n}U_{pp}\; \det \left({\tilde M}_{(n)}\right)_{(d-5)\times(d-5)}.$$
Again the matrix $\tilde{M}_{(n)}$ completely is similar to $M_{(n)}$ 
without the ($p+1$)'th row and ($p+1$)'th column.
Adding all these together we receive the following 
satisfactory limit for the prefactor (4.2)
\begin{equation}
\frac{T_p}{2}\lim_{U_{pp}\rightarrow \infty}\frac{\prod_{n=1}^{\infty}\left[\det
M_{(n)}\right]^{-1}}{\sqrt{\det{(U/4\pi \alpha')}}}~\longrightarrow
\frac{T_{p-1}}{2}\frac{\prod_{n=1}^{\infty}\left[\det
{\tilde M}_{(n)}\right]^{-1}}{\sqrt{\det{({\tilde U}/4\pi \alpha')}}}\;.
\label{infty}
\end{equation}
Note that for accomplishing this limit we used the regulation formula 
$\prod^\infty_{n=1}(na) \longrightarrow \sqrt{2\pi/a}$, and also 
we introduced the prominent relation between the tensions of a D$p$-brane and 
a D$(p-1)$-brane, i.e. $T_{p-1} = 2\pi \sqrt{\alpha'} T_p$.
The Eq. (4.3) clarifies that the total prefactor of the boundary state 
does not resist against the collapse of the brane. 

Now we demonstrate that the matrix $S_{(n)\lambda\lambda'}$
also respect to the dimensional reduction of the D$p$-brane.
To investigate this, for simplicity we suppose that the velocity has 
only one component along the $x^{p+1}$-direction.
In this case, after tachyon condensation all elements 
of $(p+1)$'th row and $(p+1)$'th column of the matrix 
$S_{(n)\lambda\lambda'}$ vanish,
except the element $S_{(n)pp}$ which tends to $-1$. However,
because of the velocity and background fields 
elements of the $(p+2)$'th row and $(p+2)$'th column remain nonzero.
We deduce that this part of the boundary state 
also does not prevent elimination 
of the $x^p$-direction of the D$p$-brane. 

For example, the matrix $S_{(n)}$ for a fractional D2-brane, parallel to
the $x^1x^2$-plane with the velocity $v$ along the $x^3$-direction,
at the infrared fixed point $U_{22}\rightarrow \infty$ 
possesses the following feature  
\begin{eqnarray}
\lim_{U_{22}\rightarrow\infty}S_{(n)} &=&
\left(
\begin{array}{cc}
\left(\Gamma_{(n)}\right)_{4\times 4} & 0 \\
0 & -\mathbf{1}_{(d-8)\times (d-8)}
\end{array}\right)\;,
\nonumber\\
\Gamma_{(n)}
&=&\left(
\begin{array}{cccc}
\Gamma_{(n)0}^{~~~0} &\Gamma_{(n)1}^{~~~0} &0 &\Gamma_{(n)3}^{~~~0} \\
\Gamma_{(n)0}^{~~~1}&\Gamma_{(n)1}^{~~~1}&0&\Gamma_{(n)3}^{~~~1}\\
0&0&-1&0\\
\Gamma_{(n)0}^{~~~3}&\Gamma_{(n)1}^{~~~3}&0&\Gamma_{(n)3}^{~~~3}
\end{array}\right)\;,
\end{eqnarray}
where the matrix elements are given by 
\begin{eqnarray}
\Gamma_{(n)0}^{~~~0}&=&\frac{\gamma^2(1+v^2)(1+\frac{1}{2n}U_{11})+
\gamma^2E^2_1}{1+\frac{1}{2n}U_{11}- \gamma^2E^2_1} \;,
\nonumber\\\nonumber\\
\Gamma_{(n)1}^{~~~0}&=&-\frac{2\gamma^2 E_1}
{1+\frac{1}{2n}U_{11}-\gamma^2 E^2_1} \;,
\nonumber\\\nonumber\\
\Gamma_{(n)0}^{~~~1}&=&-\frac{ 2\gamma^2 E_1}
{1+\frac{1}{2n}U_{11}- \gamma^2 E^2_1} \;,
\nonumber\\\nonumber\\
\Gamma_{(n)1}^{~~~1}&=&\frac{
1-\frac{1}{2n}U_{11}+ \gamma^2 E^2_1}
{1+\frac{1}{2n}U_{11}- \gamma^2 E^2_1} \;,
\nonumber\\\nonumber\\
\Gamma_{(n)3}^{~~~0}&=&-\frac{2\gamma^2 v(1+\frac{1}{2n}U_{11})}
{1+\frac{1}{2n}U_{11}- \gamma^2 E^2_1} \;,
\nonumber\\\nonumber\\
\Gamma_{(n)3}^{~~~1}&=&\frac{2\gamma^2 v E_1}
{1+\frac{1}{2n}U_{11}- \gamma^2 E^2_1} \;,
\nonumber\\\nonumber\\
\Gamma_{(n)0}^{~~~3}&=&\frac{\gamma^2 v\left[(1+\frac{1}{2n}U_{11})+
2\gamma^2 E^2_1 \right]}
{1+\frac{1}{2n}U_{11}- \gamma^2 E^2_1} \;,
\nonumber\\\nonumber\\
\Gamma_{(n)1}^{~~~3}&=&-\frac{2\gamma^2 v E_1}
{1+\frac{1}{2n}U_{11}- \gamma^2 E^2_1} \;,
\nonumber\\\nonumber\\
\Gamma_{(n)3}^{~~~3}&=&\frac{-\gamma^2 (1+v^2)(1+\frac{1}{2n}U_{11})+
\gamma^2E^2_1}{1+\frac{1}{2n}U_{11}- \gamma^2E^2_1} \;.
\nonumber
\end{eqnarray}
The electric field component is defined by $E_1 = \mathcal{F}_{01}$.
In the static case, i.e. $v= 0$, the matrix $\Gamma_{(n)}$ find 
the conventional feature, that is, the elements of its last row and 
last column, except $\Gamma_{(n)33}$, vanish, 
and the element $\Gamma_{(n)33}$ tends to $-1$. 
\section{Conclusions}

In this article we constructed the boundary state of a bosonic closed
string, emitted (absorbed) by a moving fractional D$p$-brane in the
orbifoldized spacetime 
$\mathbb{R}^{1, d-5} \times\mathbb{C}^{2}/\mathbb{Z}_{2}$ in 
the presence of the 
Kalb-Ramond field, a $U(1)$ gauge potential
and the open string tachyon field.
The boundary state equations reveal that in the brane volume 
the tachyon field induces an exotic potential on the center-of-mass 
of the closed string.

The interaction amplitude of two parallel moving fractional branes 
with the same dimension, in the presence of various background fields, 
was acquired. The variety of the adjustable parameters,
i.e. the background fields, velocities, 
the spacetime and branes dimensions, and the orbifoldized
directions, elaborates a generalized amplitude and 
an adjustable strength for the branes interaction.

For the large distances of the branes
the behavior of the interaction amplitude was studied. 
We observed that for the critical dimension $d=26$, in the large times 
the contribution of the mediated massless states quickly vanishes.
This is purely an effect of the orbifold projection.
In the special non-critical dimension, i.e. $d =28$, the contribution of 
the massless states reduces to the conventional 
case, i.e. in this dimension we receive a long-range force.
In fact, for each number of the orbifoldized directions one
can demonstrate that the damping of the long-range force is
compensated by a specific dimension of the non-critical spacetime,
while for the other dimensions the long-range 
force is removed. That is, for some dimensions it is drastically 
quenched, while for the other dimensions it is divergent.

At the end we specified effects of the tachyon condensation phenomenon on 
a moving fractional D$p$-brane with various background fields via 
its corresponding boundary state. 
We observed that advent of the fractionality, 
transverse motion and background fields  
cannot protect the brane against collapse and dimensional reduction.

ý  ý
\begin{thebibliography}{99}
\bibitem{bon} E. Cremmer, J. Scherk, Nucl. Phys. \textbf{B
 50}(1972) 222.
\bibitem{bon2} L. Clavelli, J. Shapiro,  Nucl. Phys. \textbf{B 57} (1973) 490.
\bibitem{hus} F. Hussain, R. Iengo, C. Nunez, 
Nucl. Phys. \textbf{B 497} (1997) 205.
\bibitem{bergf}O. Bergman, M. Gaberdiel, G. Lifschytz, 
Nucl. Phys. \textbf{B 509} (1998) 194.
\bibitem{pol1}J. Polchinski,  Phys. Rev. Lett. \textbf{75} (1995) 4724.
\bibitem{pol0} J. Polchinski, S. Chaudhuri, C. V. Johnson, ``\textit{Notes
on
D-branes}'',\href{http://lanl.arxiv.org/abs/hep-th/9602052}{hep-th/9602052};
 J. Polchinski, ``\textit{TASI lectures on D-branes}'',
 \href{http://lanl.arxiv.org/abs/hep-th/9611050}{hep-th/9611050}.
\bibitem{bach}C. Bachas, M. Porrati, Phys. Lett. \textbf{B 296} (1992) 77.
\bibitem{cal0}C. G. Callan, C. Lovelace, C. R. Nappi, S. A. Yost,
Nucl. Phys. \textbf{B 288} (1987) 525;  Nucl. Phys. \textbf{B 293} (1987).
\bibitem{green1}M. B. Green, P. Wai,  Nucl. Phys.
\textbf{B 431} (1994) 131.
\bibitem{greenl} M. B. Green, M.
Gutperle, Nucl. Phys. \textbf{B 476} (1996) 484-514.
\bibitem{green2}M. Billo, D. Cangemi, P. Di Vecchia, 
Phys. Lett. \textbf{B 400} (1997) 63-70.
\bibitem{khor} J. Khoury, H. Verlinde, 
Adv. Theor. Math. Phys. 3 (1999) 1893-1908.
\bibitem{fra} M. Frau, I. Pesando, S. Sciuto, A. Lerda, R. Russo, 
Phys. Lett. \textbf{B 400} 
(1997) 52.
\bibitem{calli} C. G. Callan, I. R. Klebanov, 
Nucl. Phys. \textbf{B 465} (1996) 473.
\bibitem{guk} S. Gukov, I. R. Klebanov, A. M. Polyakov, 
Phys. Lett. \textbf{B 423} (1998) 64.
\bibitem{li}M. Li, Nucl. Phys. \textbf{B 460} (1996) 351.
\bibitem{kit} T. Kitao, N. Ohta, J. G. Zhou, 
Phys. Lett. \textbf{B 428} (1998) 68.
\bibitem{bach0}C. Bachas, 
Phys. Lett. \textbf{B 374} (1996) 37.
\bibitem{doc1}
H. Arfaei and D. Kamani, Phys. Lett. \textbf{B 452} (1999) 54,
arXiv:hep-th/9909167;
Nucl. Phys. \textbf{B 561} (1999) 57-76, arXiv:hep-th/9911146; 
Phys. Lett. {\bf B 475} (2000) 39, arXiv:hep-th/9909079; 
D. Kamani, Annals of Physics {\bf 354} (2015) 394-400, arXiv:1501.02453;
Int.J.Theor.Phys.47:1533-1541,2008, arXiv:hep-th/0611339; 
Mod. Phys. Lett. {\bf A 15} (2000) 1655, arXiv:hep-th/9910043;
Phys. Lett. {\bf B 487} (2000) 187–191, arXiv:hep-th/0010019;
Nucl. Phys. {\bf B 601} (2001) 149, arXiv:hep-th/0104089;
Phys. Lett. {\bf B 487} (2000) 187, arXiv:hep-th/0010019;
F. Safarzadeh-Maleki and D. Kamani, Phys. Rev. 
\textbf{D 89}, 026006 (2014), arXiv:1312.5489;
Phys. Rev. \textbf{D 90}, 107902 (2014), arXiv:1410.4948;
J. Exp. Theor. Phys. {\bf 119} (2014) 677, arXiv:1406.2667 [hep-th];  
Z. Rezaei and D. Kamani, J. Exp. Theor. Phys. {\bf 113} (2011) 956,
arXiv:1106.2097 [hep-th];
J. Exp. Theor. Phys. {\bf 114} (2012) 234, arXiv:1107.1183 [hep-th];
E. Maghsoodi and D. Kamani, 
{\it Fractional-wrapped branes with rotation, linear motion
and background fields}, Nucl. Phys. B (2017), 
http://dx.doi.org/10.1016/j.nuclphysb.2017.07.009
\bibitem{frac0} 
M. R. Douglas, JHEP \textbf{9707} (1997) 004.
\bibitem{dog1} D.
Diaconescu, M. R. Douglas, J. Gomis,  JHEP \textbf{9802} (1998) 013.
\bibitem{frac1} C. V. Johnson, R. C. Myers, 
Phys. Rev.  \textbf{D 55} (1997) 6382-6393.
\bibitem{frac2} M. Frau, A. Liccardo, R. Musto,  
Nucl. Phys. \textbf{B 602} (2001) 39-60.
\bibitem{frac3} M. Bertolini, P. Di Vecchia, M. Frau, A. Lerda, R.
Marotta,  Nucl. Phys. \textbf{B 621} (2002)
157.
\bibitem{dive}P. Di Vecchia, A. Liccardo, R. Marotta, F. Pezzella,
 JHEP \textbf{0306} (2003) 007.
\bibitem{cn1} J. Polchinski, ``String Theory'', (Cambridge University Press, 
Cambridge, (1998) Volume I and II; C. V. Johnson, ``D-Branes'', (
Cambridge University Press, Cambridge, (2003).
\bibitem{cn2}A. Sen, Int. J. Mod. Phys. \textbf{A 14} (1999) 4061; Int. J. 
Mod. Phys. \textbf{A 20} (2005) 5513; JHEP \textbf{9808} (1998) 010; JHEP
\textbf{9808} (1998) 012; JHEP \textbf{9812} (1998) 021; JHEP \textbf{9809} 
(1998) 023; JHEP \textbf{9910} (1999) 008; JHEP \textbf{9912} (1999) 027; 
\bibitem{cn3} M. Frau, L. Gallot, A. Lerda, P. Strigazzi, Nucl. Phys. 
\textbf{B 564} (2000) 60.
\bibitem{cn4}D. Kutasov, M. Marino, G. Moore, JHEP \textbf{0010} (2000) 79.
\bibitem{cn5}P. Kraus, F. Larsen, Phys. Rev. \textbf{D 63} (2001) 106004.
\bibitem{cn6}E. Witten, Phys. Rev. \textbf{D 47} (1993) 3405; Phys. Rev.
\textbf{D 46} (1992) 5467. JHEP \textbf{9812} (1998) 019; Nucl. Phys.
\textbf{B 268} (1986) 253.
\bibitem{cn7} O. Bergman, M. R. Gaberdiel, Phys. Lett. \textbf{B 441}
(1998) 133.
\bibitem{cn8}A. Sen, JHEP \textbf{0204} (2002) -48; 
JHEP \textbf{0207} (2002) 065.
\bibitem{cn9}F. Larsen, A. Naqvi, S. Terashima, JHEP \textbf{0302} (2003) 039.
\bibitem{cn10} T. Okuda, S. Sugimoto, Nucl. Phys. \textbf{B 647} (2002) 101.
\bibitem{cn11}M. Naka, T. Takayanagi, T. Uesugi, JHEP \textbf{0006} 007.
\bibitem{cal1}C. G. Callan, C. Lovelace, C. R. Nappi, S. A. Yost, Nucl. Phys. 
\textbf{B 308} (1988) 221-284.
\bibitem{cal2}C. G. Callan, C. Lovelace, C. R. Nappi, S. A. Yost,
Nucl. Phys. \textbf{B 293} (1987) 83-113.

ý\end{thebibliography}
\end{document}